# Ehrenfest's Theorem for the Dirac Equation in Noncommutative Phase-Space

Ilyas Haouam[1]

**Abstract:** In this article, we investigate Ehrenfest's theorem from the Dirac equation in a noncommutative phase-space where we calculate the time derivative of the position and the kinetic momentum operators for Dirac particles in interaction with electromagnetic field and within a noncommutative setting. This allows examining the effect of the phase-space noncommutativity on Ehrenfest's theorem. Knowing that with both the linear Bopp-Shift and $\star$product, the noncommutativity is inserted.

**Keywords:** Dirac equation; noncommutative quantum mechanics; classical limit; Ehrenfest's theorem

**Mathematics Subject Classification:** 00A79 ; 35Q40 and 35Q72

## 1  Introduction

The Dirac equation is a relativistic quantum mechanical equation that describes the conduct of massive particles with spin-1/2, specifically known as Dirac particles, which include electrons. The classical limit (CL) of the Dirac equation can be explored when quantum mechanics' influences are disregarded, allowing us to describe the system's conduct using classical physics. Quantum effects such as entanglement, superposition and interference should vanish at the macroscopic scale in the CL, and this demise is not easy to explain. However, in this scenario, the system behaves according to classical laws of physics. The CL is ordinarily expressed in terms of the limit of a vanishing Planck's constant, i.e., $\hbar \rightarrow 0$ as scaled with the system's action. In this case, Hamilton's principle takes its classical expression and all the operator's commute. To explore the CL of the Dirac equation, one can begin by examining the equation's solutions in the limit of extensive distances and durations or in the limit of large energies and momenta, where the influences of quantum mechanics become negligible [36]. In specific terms, the CL emerges when the system possesses a large quantum number, undergoes significant interactions with its surroundings, or when its de Broglie wavelength becomes significantly smaller compared to other pertinent length measurements. In this scenario, the findings predicted by quantum mechanics must converge to those predicted by classical mechanics as reaching the macroscopic scale. Put differently, one can attain the CL of a quantum mechanical system by contrasting the predictions of the quantum mechanical system with those of classical physics in situations where quantum effects are minimal. If the forecasts from the Dirac equation align with those of classical physics, it implies that the CL has been attained. The frequently cited correspondence principle is perhaps the Bohr correspondence principle [2], asserts that in the limit of large quantum numbers, a quantum system exhibits behavior akin to the corresponding classical system. Numerous studies exploring the alignment between quantum and classical aspects have been conducted in the literature [26], [10], [3], [31], [25], [1], [27], [28], [29], [24].

---

[1] Laboratoire de Physique Mathématique et de Physique Subatomique (LPMPS), Université Frères Mentouri, Constantine 25000, Algeria, Email : ilyashaouam@live.fr ; ilyas.haouam@tmp.umc.edu.dz

Ehrenfest's theorem is frequently employed when studying the CL of quantum mechanical systems [5]. This theorem serves as a potent instrument for comprehending the behavior of such systems. By its application, we can see the manner in which quantum mechanical influences dissipate, giving rise to classical dynamics [6]. Certainly, Ehrenfest's theorem establishes a connection between the evolution of expected values of observables and classical equations of motion. Within the framework of the Dirac equation, this theorem remains valid in the CL, where quantum effects are insignificantly small, causing the equation to simplify to its classical counterpart. In this regime, the anticipated values of observables mirror the behavior dictated by classical equations of motion, derived from the system's Hamiltonian. Consequently, it can be asserted that Ehrenfest's theorem is applicable to the CL of the Dirac equation. Another aspect to consider is the "semiclassical limit" of the Dirac equation, which can be attained when external potentials vary slowly, such as the electrostatic potential [34]. Likewise, there is what we call the "non-relativistic limit" of the Dirac equation [11], [12], which is often taken to be the limit in which the speed of the particle is much less than the speed of light $v \ll c$ or low energy in front of the rest energy, it allows us to neglect the relativistic effects. In this limit, the Dirac equation reduces to the Schrödinger equation, which is the equation of motion of non-relativistic quantum mechanics. Furthermore, the classical and the NR limits are related but distinct concepts, they address different aspects of the system's behavior. The CL focuses on the transition from the quantum to the classical realm, while the non-relativistic limit deals with the absence of relativistic effects and corrections. However, it's important to highlight that in numerous practical physical situations, the CL and the non-relativistic limit can align, leading to similar descriptions of the system's conduct.

On the other side, studies and inquiries into noncommutative (NC) geometry have emerged as a subject of significant intrigue in the past few decades. NC geometry exerts substantial influence across various domains of high-energy and contemporary physics including quantum physics [13], [30], [14], string theories, [33], standard model [32], quantum field theory [15], [35], cosmology, black holes, and gravity [9], [7]. Research has demonstrated the significance of NC phase-space (NCPS) in comprehending phenomena occurring at extremely short distances and within the high-energy ranges. It is worth noting also that various forms of phase-space noncommutativity have been explored in the literature as part of a comprehensive examination of this subject. For an overview of this subject, check [16], [17], [18], [19]. The passage from the commutative setting to the NC one (or vice versa) is achieved using four essential methods namely, (i) Bopp-shift transformations. (ii) Moyal–Weyl product (⋆product) instead of ordinary product in the actions and functions of the physical systems. (iii) Seiberg–Witten maps. (iv) Weyl–Wigner maps, which means using the ordinary product with Weyl operators. As has been extensively discussed in the literature (see, e.g., [13], [4], [20], [8], [21], [22], [23], and references therein). However, here in this study, our approach hinges on the combination of the ⋆product and Bopp-shift techniques.

Inspired by endeavors to comprehend string theory, quantum gravity approaches with a focus on gravitational stability, and models of black holes, and driven by the ambition to develop a quantum formulation of the Einstein equation within the realm of NC geometry, we have chosen to investigate the problem at hand within the framework of NC geometry. Furthermore, we believe that exploring the CL of the Dirac equation can provide valuable insights into classical phenomenological implications. That is why we focus on investigating the CL through Ehrenfest's theorem from the Dirac equation in the NC setting. This paper is outlined as follows: In Section 2, the NC geometry is briefly reviewed. In Section 3, we study the CL of the NC Dirac equation whereas in Sub-Section 3.1, a deformed Dirac equation is obtained. In Sub-Section 3.2, we investigate Ehrenfest's theorem from the Dirac equation within the NC framework, and then Section 4 is left for the conclusion and remarks.

## 2  Brief Overview on the Noncommutative Phase-Space

The NC geometry offers intriguing characteristics within the realm of quantum mechanics, and in order to explore these NC effects, we've adapted quantum mechanics to incorporate noncommutativity. Consequently, in this segment of our investigation, we provide a concise overview of the specific NC geometry required for our calculations. However, when dealing with extremely minuscule scales, such as those found in string theory, the position coordinates no longer commute, nor do the momenta. This implies that the order of operators matters during calculations. So, for two different operators, e.g., $\hat{A}$ and $\hat{B}$, one has, in general, $\hat{A} * \hat{B} \neq \hat{B} * \hat{A}$. Now, we consider the operators for coordinates and momentum in the NC setting as $x_i^{nc}$ and $p_i^{nc}$ respectively, then, in a d-dimensional NCPS, the altered commutation relations take the following form:

$$\begin{aligned} \left[x_\mu^{nc}, x_\nu^{nc}\right] &= i\Theta_{\mu\nu}, \\ \left[p_\mu^{nc}, p_\nu^{nc}\right] &= i\eta_{\mu\nu}, \quad (\mu, \nu = 1, \ldots d), \\ \left[x_\mu^{nc}, p_\nu^{nc}\right] &= i\widetilde{\hbar}\delta_{\mu\nu} \end{aligned} \quad (2.1)$$

with $\Theta_{\mu\nu}$, $\eta_{\mu\nu}$ stand for constant antisymmetric $d \times d$ matrices and $\delta_{\mu\nu}$ is the identity matrix. The effective Planck constant (deformed $\hbar$) is given by

$$\widetilde{\hbar} = \hbar(AB + \xi), \text{ with } \xi = \frac{Tr[\Theta\eta]}{4AB\hbar^2}, \quad (2.2)$$

where $\xi \ll 1$ is the consistency condition in quantum mechanics, and is expected to be typically satisfied when the NC parameters are of the 2nd order. Now, $x_i^{nc}$ and $p_i^{nc}$ can be presented in terms of the commutative coordinates $x_i$ and $p_i$ in conventional quantum mechanics via the following Bopp-shift [20]

$$x_\mu^{nc} = Ax_\mu - \frac{\Theta_{\mu\nu}}{2A\hbar}p_\nu \; ; \; p_\mu^{nc} = Bp_\mu + \frac{\eta_{\mu\nu}}{2B\hbar}x_\nu, \quad (2.3)$$

where $A = 1 - \frac{\Theta\eta}{8\hbar^2}$ and $B = \frac{1}{A}$, which are scaling constants and usually are considered $A = B = 1$ to the 1st order of $\Theta$ and $\eta$. As a consequence, the equations (2.3) and (2.2) become

$$\begin{aligned} x_\mu^{nc} &= x_\mu - \frac{\Theta_{\mu\nu}}{2\hbar}p_\nu, \\ p_\mu^{nc} &= p_\mu + \frac{\eta_{\mu\nu}}{2\hbar}x_\nu, \end{aligned} \quad \widetilde{\hbar} = \hbar(1 + \xi|_{A=B=1}). \quad (2.4)$$

As long as the system under consideration is three-dimensional, we have the following NC algebra:

$$\begin{aligned} \left[x_j^{nc}, x_k^{nc}\right] &= \frac{i}{2}\epsilon_{jkl}\Theta_l, \\ \left[p_j^{nc}, p_k^{nc}\right] &= \frac{i}{2}\epsilon_{jkl}\eta_l, \quad (j, k, l = 1, 2, 3), \\ \left[x_j^{nc}, p_k^{nc}\right] &= i\hbar\left(1 + \frac{\Theta\eta}{4\hbar^2}\right)\delta_{jk}, \end{aligned} \quad (2.5)$$

with $\Theta_l = (0,0,\Theta)$, $\eta_l = (0,0,\eta)$ where $\Theta$, $\eta$ are real-valued NC parameters with $[\Theta] = L^2$ and $[\eta] = M^2 L^2 T^{-2}$, and $\epsilon_{jkl}$ is the Levi-Civita permutation tensor. These NC parameters are supposed to be extremely small, consequently, we have

$$x_i^{nc} = x_i - \frac{\epsilon_{ijk}\Theta_k}{4\hbar}p_j \;\; ; \;\; p_i^{nc} = p_i + \frac{\epsilon_{ijk}\eta_k}{4\hbar}x_j. \tag{2.6}$$

In NC quantum mechanics, replacing the usual product with the ⋆product, effectively transforms our quantum mechanical system into an NC one. Furthermore, if let $\mathcal{H}(x,p)$ be the Hamiltonian operator of the standard quantum mechanics, then its corresponding Schrödinger equation in NC quantum mechanics typically be

$$\mathcal{H}(x,p) \star \psi(x,p) = E\psi(x,p). \tag{2.7}$$

The ⋆product between two arbitrary functions $\mathcal{F}(x,p)$ and $\mathcal{G}(x,p)$ from $\mathcal{R}^D$ within phase-space is defined as [13],

$$(\mathcal{F} \star \mathcal{G})(x,p) = \exp\left[\frac{i}{2}\Theta_{ab}\partial_{x_a}\partial_{x_b} + \frac{i}{2}\eta_{ab}\partial_{p_a}\partial_{p_b}\right]\mathcal{F}(x_a,p_a)\mathcal{G}(x_b,p_b) = \mathcal{F}(x,p)\,\mathcal{G}(x,p) +$$
$$\sum_{n=1}\frac{1}{n!}\left(\frac{i}{2}\right)^n \Theta^{a_1 b_1} \dots \Theta^{a_n b_n}\partial_{a_1}^x \dots \partial_{a_n}^x \mathcal{F}(x,p) \times \partial_{b_1}^x \dots \partial_{b_n}^x \mathcal{G}(x,p) +$$
$$\sum_{n=1}\frac{1}{n!}\left(\frac{i}{2}\right)^n \eta^{a_1 b_1} \dots \eta^{a_n b_n}\partial_{a_1}^p \dots \partial_{a_n}^p \mathcal{F}(x,p) \times \partial_{b_1}^p \dots \partial_{b_n}^p \mathcal{G}(x,p), \tag{2.8}$$

with $\mathcal{F}(x,p)$ and $\mathcal{G}(x,p)$, supposed to be infinitely differentiable. But once consider only the case of NC space, the ⋆product definition becomes [23],

$$(\mathcal{F} \star \mathcal{G})(x) = \exp\left[\frac{i}{2}\Theta_{ab}\partial_{x_a}\partial_{x_b}\right]\mathcal{F}(x_a)\mathcal{G}(x_b) = \mathcal{F}(x)\,\mathcal{G}(x) + \sum_{n=1}\frac{1}{n!}\left(\frac{i}{2}\right)^n \Theta^{a_1 b_1} \dots \Theta^{a_n b_n}\partial_{a_1} \dots \partial_{a_n}\mathcal{F}(x) \times$$
$$\partial_{b_1} \dots \partial_{b_n}\mathcal{G}(x). \tag{2.9}$$

For low-energy fields, the NC field theories where $E^2 \leq 1/\Theta$ at the classical level are entirely reduced to their commutative counterparts due to the ⋆product nature. However, this is only the classical result, and even at low energies, quantum corrections inevitably reveal the influences of NC parameters. Knowing that in quantum mechanics, terms involving NC parameters can always be treated as perturbations. When let $\Theta = \eta = 0$, the NC algebra simplifies to the ordinary commutative one and the canonical commutation relations become

$$\begin{aligned}[x_j,x_k] &= 0, \\ [p_j,p_k] &= 0, \quad (j,k = 1,2,3). \\ [x_i,x_i] &= i\hbar\delta_{jk},\end{aligned} \tag{2.10}$$

## 3 Classical Limit of the Noncommutative Dirac Equation

In this section, we obtain the Dirac equation in the NC framework and then use it to investigate Ehrenfest's theorem.

### 3.1 Noncommutative Dirac equation

In a commutative phase-space, the time-dependent Dirac equation in interaction with an electromagnetic four-potential $A^\mu(\vec{A},\Phi)$ is:

$$\left\{c\vec{\alpha}\cdot\left(\vec{p}-\frac{e}{c}\vec{A}(\vec{r})\right)+e\Phi(\vec{r})+\beta mc^2\right\}\psi(\vec{r},t)=E\psi(\vec{r},t), \quad (3.1)$$

where $\psi(\vec{r},t)=\left(\phi(\vec{r},t)\ \chi(\vec{r},t)\right)^T$ is the bispinor in the Dirac representation. The momentum $\vec{p}$ is given by $\vec{p}=-i\hbar\vec{\nabla}$ and the Dirac matrices $\alpha_i$ and $\beta$ satisfy the following anticommutation relations:

$$\{\alpha_i,\alpha_j\}=2\delta_{ij},\ \{\alpha_i,\beta\}=0,\ \alpha_i^2=\beta^2. \quad (3.2)$$

Note that in equation (3.1), we utilized $\frac{e}{c}\vec{A}(\vec{r})$ instead of $e\vec{A}(\vec{r})$. The factor $\frac{e}{c}$ is employed in Gaussian units, whereas in SI units, there is no such factor. In the following, we derive the NC Dirac equation through two steps, the first step is by implementing space noncommutativity while in the second one, we implement the phase noncommutativity. So, we first implement the noncommutativity in space by linking the commutative coordinates $\vec{r}$ with the NC ones $\vec{r}^{nc}$ through the $\star$product, therefore with the help of equation (2.9), we obtain

$$\mathcal{H}(\vec{r},\vec{p})\star\psi(\vec{r},t)=\left\{c\vec{\alpha}\cdot\left(\vec{p}-\frac{e}{c}\vec{A}(\vec{r})\right)+e\Phi(\vec{r})+\beta mc^2\right\}\star\psi(\vec{r},t)=i\hbar\frac{\partial}{\partial t}\psi(\vec{r},t). \quad (3.3)$$

Knowing that the electromagnetic potential adopts the form $A(\vec{r})\propto Br$, which is expressed in the symmetric gauge, where $B$ is the magnetic field, thus, the derivations in equation (2.9) firmly turned off in the first-order, consequently, equation (3.3) becomes

$$\mathcal{H}(\vec{r},\vec{p})\star\psi(\vec{r},t)=\mathcal{H}(\vec{r},\vec{p})\psi(\vec{r},t)+\frac{i}{2}\Theta_{ab}\partial_a\left\{c\vec{\alpha}\cdot\left(\vec{p}-\frac{e}{c}\vec{A}(\vec{r})\right)+e\Phi(\vec{r})+\beta mc^2\right\}\partial_b\psi(\vec{r},t)+\mathcal{O}(\Theta^2)=i\hbar\frac{\partial}{\partial t}\psi(\vec{r},t). \quad (3.4)$$

Note that $c\vec{\alpha}\cdot(\partial_a\vec{p})=\beta\partial_a(mc^2)=0$, then the equation (3.4) simply becomes

$$\mathcal{H}(\vec{r},\vec{p})\psi(\vec{r},t)-\frac{ie}{2}\Theta_{ab}\partial_a\{\vec{\alpha}\cdot\vec{A}+\Phi\}\partial_b\psi(\vec{r},t)=i\hbar\frac{\partial}{\partial t}\psi(\vec{r},t). \quad (3.5)$$

Now, we move on to implement the noncommutativity in phase in equation (3.5), thus, we relate the commutative momentum $\vec{p}$ with the NC one $\vec{p}^{nc}$ via the Bopp-shift transformation, so using equation (2.6), we obtain the phase-space NC Dirac equation:

$$\left\{c\vec{\alpha}\cdot\left(\vec{p}+\frac{1}{2\hbar}\eta_{ij}x_j-\frac{e}{c}\vec{A}\right)+e\Phi+\beta mc^2-\frac{ie}{2}\Theta_{ab}\partial_a(\vec{\alpha}\cdot\vec{A}+\Phi)\partial_b\right\}\psi=i\hbar\frac{\partial}{\partial t}\psi. \quad (3.6)$$

Following necessary minor simplifications, equation (3.6) can be expressed in a more concise form as follows [11], [12]:

$$\left\{c\vec{\alpha}\cdot\left(\vec{p}-\frac{e}{c}\vec{A}\right)+e\Phi+\beta mc^2+\frac{c}{\hbar}(\vec{\alpha}\times\vec{r})\cdot\vec{\eta}+\frac{e}{\hbar}\left(\vec{\nabla}\left(\vec{\alpha}\cdot\vec{A}-\Phi\right)\times\vec{p}\right)\cdot\vec{\Theta}\right\}\psi^{nc}=i\hbar\frac{\partial}{\partial t}\psi^{nc}, \quad (3.7)$$

where $\psi^{nc}(\vec{r},t)=\left(\phi(\Theta,\eta)\ \chi(\Theta,\eta)\right)^T$ is the wave function in the NC phase-space. Next, we move to employ the obtained deformed Dirac equation (3.7) to explore the CL through Ehrenfest's theorem.

## 3.2 Ehrenfest's theorem in noncommutative setting

Ehrenfest's theorem can be derived from the Dirac equation. It asserts that the time evolution of expected values of observables in quantum mechanics adheres to classical equations of motion. In simpler terms, it implies that the average behavior of a quantum system aligns with classical physics. It is also important to note that this theorem is applicable to all quantum systems. However, in this context, we are calculating the time derivatives of position and kinetic momentum operators for Dirac particles interacting with an electromagnetic field within a deformed framework. So, the equation of motion of an arbitrary operator $\hat{\mathcal{F}}$ is given by

$$\frac{d\hat{\mathcal{F}}}{dt} = \frac{\partial \hat{\mathcal{F}}}{\partial t} + \frac{i}{\hbar}[\hat{\mathcal{H}}, \hat{\mathcal{F}}], \qquad (3.8)$$

where $\hat{\mathcal{H}}$ is the Hamiltonian operator. Let's start with the position operator

$$\frac{d\hat{\vec{x}}}{dt} = \frac{\partial \hat{\vec{x}}}{\partial t} + \frac{i}{\hbar}[\hat{\mathcal{H}}^{nc}, \hat{\vec{x}}] = \frac{i}{\hbar}[\hat{\mathcal{H}}^{nc}, \hat{\vec{x}}], \qquad (3.9)$$

where the Hamiltonian operator from equation (3.7) is

$$\hat{\mathcal{H}}^{nc} = c\hat{\vec{\alpha}} \cdot \left(\hat{\vec{p}} - \frac{e}{c}\vec{A}\right) + e\Phi + \hat{\beta}mc^2 + \frac{c}{\hbar}(\hat{\vec{\alpha}} \times \hat{\vec{r}}) \cdot \vec{\eta} + \frac{e}{\hbar}\left(\vec{\nabla}\left(\hat{\vec{\alpha}} \cdot \vec{A} - \Phi\right) \times \hat{\vec{p}}\right) \cdot \vec{\Theta}, \qquad (3.10)$$

then the commutator in (3.9) reads

$$[\hat{\mathcal{H}}^{nc}, \hat{\vec{x}}] = c[\hat{\vec{\alpha}} \cdot \hat{\vec{p}}, \hat{\vec{x}}] - e[\hat{\vec{\alpha}} \cdot \vec{A}, \hat{\vec{x}}] + e[\Phi, \hat{\vec{x}}] + mc^2[\hat{\beta}, \hat{\vec{x}}] + \frac{c}{\hbar}[(\hat{\vec{\alpha}} \times \hat{\vec{r}}) \cdot \vec{\eta}, \hat{\vec{x}}] + \frac{e}{\hbar}[(\vec{\nabla}(\hat{\vec{\alpha}} \cdot \vec{A} - \Phi) \times \hat{\vec{p}}) \cdot \vec{\Theta}, \hat{\vec{x}}], \qquad (3.11)$$

$\Phi$ depends on spatial coordinates, thus $[\Phi, \hat{\vec{x}}] = 0$. The position operator $\hat{x}$ is diagonal concerning the spinor indices, i.e., $\hat{\vec{x}}\psi = \vec{x}\psi$ and contains no differentiation, consequently $[\hat{\beta}, \hat{\vec{x}}] = [\hat{\vec{\alpha}}, \hat{\vec{x}}] = 0$, for three arbitrary vectors $\vec{A}_1$, $\vec{A}_2$ and $\vec{A}_3$ we use the identity $[\vec{A}_1 \vec{A}_2, \vec{A}_3] = [\vec{A}_1, \vec{A}_3]\vec{A}_2 + \vec{A}_1[\vec{A}_2, \vec{A}_3]$. Then we have

$$[\hat{\vec{\alpha}} \cdot \hat{\vec{p}}, \hat{\vec{x}}] = -i\hbar\hat{\vec{\alpha}}, \qquad (3.12)$$

and

$$[\hat{\vec{\alpha}} \cdot \vec{A}, \hat{\vec{x}}] = 0, \qquad (3.13)$$

also

$$[(\hat{\vec{\alpha}} \times \hat{\vec{r}}) \cdot \vec{\eta}, \hat{\vec{x}}] = 0. \qquad (3.14)$$

Through the feature $\vec{A}_1 \cdot (\vec{A}_2 \times \vec{A}_3) = \vec{A}_2 \cdot (\vec{A}_3 \times \vec{A}_1) = \vec{A}_3 \cdot (\vec{A}_1 \times \vec{A}_2)$, we have

$$[(\vec{\nabla}(\hat{\vec{\alpha}} \cdot \vec{A} - \Phi) \times \hat{\vec{p}}) \cdot \vec{\Theta}, \hat{\vec{x}}] = \sum_{i,j}[\hat{p}_i, \hat{x}_j]\left(\vec{\Theta} \times \vec{\nabla}(\hat{\vec{\alpha}} \cdot \vec{A} - \Phi)\right)_i e_j = -i\hbar \sum_{i,j}\delta_{ij}\left(\vec{\Theta} \times \vec{\nabla}(\hat{\vec{\alpha}} \cdot \vec{A} - \Phi)\right)_i e_j, \qquad (3.15)$$

with $[\hat{p}_i, \hat{x}_j] = -i\hbar\delta_{ij}$, then we obtain

$$\frac{d\hat{\vec{x}}}{dt} = c\hat{\vec{\alpha}} - ie \sum_{i,j}\delta_{ij}\left(\vec{\Theta} \times \vec{\nabla}(\hat{\vec{\alpha}} \cdot \vec{A} - \Phi)\right)_i e_j \equiv \hat{\vec{v}} + \vec{\Lambda}, \qquad (3.16)$$

where $\hat{\vec{v}} = c\hat{\vec{\alpha}}$ and $\vec{\Lambda} = ie \sum_{i,j} \delta_{ij} \left(\vec{\Theta} \times \vec{\nabla} \left(\hat{\vec{\alpha}} \cdot \vec{A} - \Phi\right)\right)_i e_j$. Let us subsequently see how the above operator (3.16) acts on the Dirac spinor. Considering single components $\psi$, then we get

$$\frac{d\hat{\vec{x}}}{dt}\psi = \pm c\psi - ie\sum_{i,j}\delta_{ij}\left(\vec{\Theta}\times\vec{\nabla}\left(\hat{\vec{\alpha}}\cdot\vec{A}-\Phi\right)\right)_i e_j \, |_{\hat{\vec{\alpha}}\equiv\pm 1}\psi, \qquad (3.17)$$

where the eigenvalues of $\hat{\vec{\alpha}}$ are $\pm 1$. In the limit of $\Theta \to 0$, equation (3.17) turns to:

$$\frac{d\hat{\vec{x}}}{dt}\psi = \pm c\psi. \qquad (3.18)$$

This result has no classical analogy because despite the considered effects the Dirac particle is still moving at the speed of light as appears in the first term of equations (3.17) and (3.18) (case when $\vec{\Lambda} = 0$).

Now, the equation of motion for the kinetic momentum $\hat{\vec{D}} = \hat{\vec{p}} - \frac{e}{c}\vec{A}$ is

$$\frac{d\hat{\vec{D}}}{dt} = \frac{\partial \hat{\vec{D}}}{\partial t} + \frac{i}{\hbar}\left[\mathcal{H}^{nc}, \hat{\vec{D}}\right] = -\frac{e}{c}\frac{\partial \vec{A}}{\partial t} + \frac{i}{\hbar}\left[\mathcal{H}^{nc}, \hat{\vec{D}}\right]. \qquad (3.19)$$

Then, the commutator is given by

$$\left[\mathcal{H}^{nc}, \hat{\vec{D}}\right] = \left[\mathcal{H}^{nc}, \hat{\vec{p}}\right] - \frac{e}{c}\left[\mathcal{H}^{nc}, \vec{A}\right]. \qquad (3.20)$$

First, we calculate the first commutator in (3.20)

$$\left[\mathcal{H}^{nc}, \hat{\vec{p}}\right] = c[\hat{\vec{\alpha}}\cdot\hat{\vec{p}}, \hat{\vec{p}}] - e[\hat{\vec{\alpha}}\cdot\vec{A}, \hat{\vec{p}}] + e[\Phi, \hat{\vec{p}}] + mc^2[\hat{\beta}, \hat{\vec{p}}] + \frac{c}{\hbar}[(\hat{\vec{\alpha}}\times\hat{\vec{r}})\cdot\vec{\eta}, \hat{\vec{p}}] + \frac{e}{\hbar}[(\vec{\nabla}(\hat{\vec{\alpha}}\cdot\vec{A} - \Phi)\times\hat{\vec{p}})\cdot\vec{\Theta}, \hat{\vec{p}}], \qquad (3.21)$$

with $[\hat{\beta}, \hat{\vec{p}}] = [\hat{\vec{\alpha}}, \hat{\vec{p}}] = 0$ because $\hat{\beta}$ and $\hat{\vec{\alpha}}$ do not depend on space coordinates. Furthermore, we have

$$e[\Phi, \hat{\vec{p}}] = i\hbar e[\vec{\nabla}, \Phi] = i\hbar e(\vec{\nabla}\Phi - \Phi\vec{\nabla}), \qquad (3.22)$$

then making use of this commutator to get

$$e[\Phi, \hat{\vec{p}}]\psi = i\hbar e(\vec{\nabla}\Phi - \Phi\vec{\nabla})\psi = i\hbar e(\vec{\nabla}\Phi)\psi, \qquad (3.23)$$

and

$$c[\hat{\vec{\alpha}}\cdot\hat{\vec{p}}, \hat{\vec{p}}] = 0, \qquad (3.24)$$

also

$$-e[\hat{\vec{\alpha}}\cdot\vec{A}, \hat{\vec{p}}]\psi = -e\sum_{i,j}\hat{\alpha}_i[A_i, \hat{p}_j]e_j\,\psi = -i\hbar e\sum_{i,j}\hat{\alpha}_i[A_i, \nabla_j]e_j\,\psi = -i\hbar e\sum_{i,j}\hat{\alpha}_i(\nabla_j A_i)e_j\,\psi. \qquad (3.25)$$

By using the feature above the equation (3.15), we get

$$\frac{c}{\hbar}[(\hat{\vec{\alpha}}\times\hat{\vec{r}})\cdot\vec{\eta}, \hat{\vec{p}}] = \frac{c}{\hbar}\sum_{i,j}\left[(\hat{\vec{\alpha}}\times\vec{\eta})_i \hat{r}_i, \hat{p}_j\right]e_j = -ic\sum_{i,j}[\hat{r}_i, \nabla_j](\vec{\eta}\times\hat{\vec{\alpha}})_i e_j, \qquad (3.26)$$

with

$$\frac{e}{\hbar}\left[\left(\vec{\nabla}\left(\hat{\vec{\alpha}}\cdot\vec{A}-\Phi\right)\times\hat{\vec{p}}\right)\cdot\vec{\Theta},\hat{\vec{p}}\right] = \frac{e}{\hbar}\left[\left(\vec{\Theta}\times\vec{\nabla}\left(\hat{\vec{\alpha}}\cdot\vec{A}-\Phi\right)\times\hat{\vec{p}}\right)\cdot\hat{\vec{p}},\hat{\vec{p}}\right] = \frac{e}{\hbar}\sum_{i,j}\left[\left(\vec{\Theta}\times\vec{\nabla}\left(\hat{\vec{\alpha}}\cdot\vec{A}-\Phi\right)\right)_i,\hat{p}_j\right]\hat{p}_i e_j =$$
$$-ie\sum_{i,j}\left[\left(\vec{\Theta}\times\vec{\nabla}\left(\hat{\vec{\alpha}}\cdot\vec{A}-\Phi\right)\right)_i,\nabla_j\right]\nabla_i e_j. \qquad (3.27)$$

Now, for the second commutator in equation (3.20), we have

$$[\mathcal{H}^{nc},\vec{A}] = c[\hat{\vec{\alpha}}\cdot\hat{\vec{p}},\vec{A}] - e[\hat{\vec{\alpha}}\cdot\vec{A},\vec{A}] + e[\Phi,\vec{A}] + mc^2[\hat{\beta},\vec{A}] + \frac{c}{\hbar}[(\hat{\vec{\alpha}}\times\hat{\vec{r}})\cdot\vec{\eta},\vec{A}] + \frac{e}{\hbar}[(\vec{\nabla}(\hat{\vec{\alpha}}\cdot\vec{A}-\Phi)\times\hat{\vec{p}})\cdot\vec{\Theta},\vec{A}]. \qquad (3.28)$$

Thereafter we move to calculate each commutator in equation (3.28), thus we start with

$$c[\hat{\vec{\alpha}}\cdot\hat{\vec{p}},\vec{A}]\psi = c\sum_{i,j}\hat{\alpha}_i[\hat{p}_i,A_j]e_j\psi = -i\hbar c\sum_{i,j}\hat{\alpha}_i(\nabla_i A_j)e_j\psi, \qquad (3.29)$$

where the gradient in equations (3.25) and (3.29) acts on $\vec{A}$ only.
Knowing that $[\hat{\beta},\vec{A}] = [\Phi,\vec{A}] = [\hat{\vec{\alpha}},\vec{A}] = 0; \frac{c}{\hbar}\sum_{i,j}\left[(\hat{\vec{\alpha}}\times\hat{\vec{r}})_i,A_j\right]\eta_i e_j = 0$, and

$$\frac{e}{\hbar}\left[\left(\vec{\nabla}\left(\hat{\vec{\alpha}}\cdot\vec{A}-\Phi\right)\times\hat{\vec{p}}\right)\cdot\vec{\Theta},\vec{A}\right] = \frac{e}{\hbar}\left[\left(\vec{\Theta}\times\vec{\nabla}\left(\hat{\vec{\alpha}}\cdot\vec{A}-\Phi\right)\right)\cdot\hat{\vec{p}},\vec{A}\right] = \frac{e}{\hbar}\sum_{i,j}\left(\vec{\Theta}\times\vec{\nabla}\left(\hat{\vec{\alpha}}\cdot\vec{A}-\Phi\right)\right)_i[\hat{p}_i,A_j]e_j =$$
$$-ie\sum_{i,j}\left(\vec{\Theta}\times\vec{\nabla}\left(\hat{\vec{\alpha}}\cdot\vec{A}-\Phi\right)\right)_i(\nabla_i A_j)e_j. \qquad (3.30)$$

Now in total, we have

$$\frac{d\hat{\vec{D}}}{dt} = -e\left(\frac{1}{c}\frac{\partial\vec{A}}{\partial t}+\vec{\nabla}\Phi\right) + \frac{e}{c}\sum_{i,j}(c\hat{\alpha}_i)(\nabla_j A_i - \nabla_i A_j)e_j + \frac{c}{\hbar}\sum_{i,j}[\hat{r}_i,\nabla_j](\vec{\eta}\times\hat{\vec{\alpha}})_i e_j + \frac{e}{\hbar}\sum_{i,j}\left[\left(\vec{\Theta}\times\vec{\nabla}\left(\hat{\vec{\alpha}}\cdot\vec{A}-\Phi\right)\right)_i,\nabla_j\right]\nabla_i e_j - \frac{e^2}{c\hbar}\sum_{i,j}\left(\vec{\Theta}\times\vec{\nabla}\left(\hat{\vec{\alpha}}\cdot\vec{A}-\Phi\right)\right)_i(\nabla_i A_j)e_j. \qquad (3.31)$$

After some simplifications we get

$$\frac{d\hat{\vec{D}}}{dt} = e\vec{E} + \frac{e}{c}\vec{v}\times\text{curl}\vec{A} + \frac{e}{\hbar}\sum_{i,j}\left[\left(\vec{\Theta}\times\vec{\nabla}\left(\hat{\vec{\alpha}}\cdot\vec{A}-\Phi\right)\right)_i,\nabla_j\right]\nabla_i e_j + \frac{c}{\hbar}\sum_{i,j}[\hat{r}_i,\nabla_j](\vec{\eta}\times\hat{\vec{\alpha}})_i e_j - \frac{e^2}{c\hbar}\sum_{i,j}\left(\vec{\Theta}\times\vec{\nabla}\left(\hat{\vec{\alpha}}\cdot\vec{A}-\Phi\right)\right)_i(\nabla_i A_j)e_j, \qquad (3.32)$$

where $-\vec{E} = \frac{1}{c}\frac{\partial\vec{A}}{\partial t}+\vec{\nabla}\Phi$ and $\vec{v}\times\text{curl}\vec{A} = \sum_{i,j}\hat{v}_i(\nabla_j A_i - \nabla_i A_j)e_j$.
With $\vec{v}\times\text{curl}\vec{A} = \vec{v}\times\vec{B}$, then we have

$$\frac{d\hat{\vec{D}}}{dt} = e\vec{E} + \frac{e}{c}\vec{v}\times\vec{B} + \frac{e}{\hbar}\sum_{i,j}\left[\left(\vec{\Theta}\times\vec{\nabla}\left(\hat{\vec{\alpha}}\cdot\vec{A}-\Phi\right)\right)_i,\nabla_j\right]\nabla_i e_j + \frac{c}{\hbar}\sum_{i,j}[\hat{r}_i,\nabla_j](\vec{\eta}\times\hat{\vec{\alpha}})_i e_j - \frac{e^2}{c\hbar}\sum_{i,j}\left(\vec{\Theta}\times\vec{\nabla}\left(\hat{\vec{\alpha}}\cdot\vec{A}-\Phi\right)\right)_i(\nabla_i A_j)e_j, \qquad (3.33)$$

which is a deformed Lorentz force in the classical case, and in the limits of $\Theta \to 0$ and $\eta \to 0$, we have

$$\frac{d\hat{\vec{D}}}{dt} = e\vec{E} + \frac{e}{c}\vec{v}\times\vec{B}, \qquad (3.34)$$

which is the Lorentz force in the classical case. Now we move to discuss the results, so we find that $\hat{\vec{x}}$ does not satisfy classical equations of motion, but a classical equation of motion can be set up for the operator $\hat{\vec{D}}$. However, at least equation (3.34) seems to coincide formally with the corresponding classical equation, although we always have to bear in mind that any expectation values of (3.34) are not very useful due to the Zitterbewegung, reduced in the velocity. At most, the projection of the even contributions of (3.34) produces results, which are relevant within the scope of a classical single-particle description. We now return to equation (3.32), which shows the effect of the phase-space noncommutativity on the form of the Lorentz force, which has been deformed by these considerations, also equation (3.17) shows the effect of noncommutativity on the obtained velocity. However, we find that the used conditions and considerations affected Ehrenfest's theorem.

## 3 Conclusion and remarks

The classical and non-relativistic limits are essential in quantum mechanics; thus, it is very interesting to investigate fundamental phenomena in these limits. So, studying the CL of the Dirac equation is important for understanding the behavior of Dirac particles in a variety of contexts, from condensed matter systems to particle physics and cosmology. Here are some motivations for studying the CL of the Dirac equation: The CL of the Dirac equation is important for understanding how classical behavior emerges from quantum dynamics. Also, the limits of measurement in quantum mechanics. And the conduct of spin-1/2 particles in non-relativistic systems. Then, studying the CL of the Dirac equation can help us understand how the relativistic conduct of particles reduces to classical mechanics.

Besides, as applications to quantum field theory, the Dirac equation is the framework for describing the conduct of particles at the most fundamental level. Thus, studying the CL of the Dirac equation can provide insight into the conduct of particles in quantum field theory, particle physics, and cosmology. We may add some advantages and applications, so in the CL, the trajectories of particles become well-defined and can be described using classical mechanics, and understanding the CL of the Dirac equation can provide insight into how classical trajectories emerge from the underlying quantum dynamics. The Dirac equation is used to describe the conduct of relativistic particles in condensed matter systems, such as graphene. Thus, investigating the CL of the Dirac equation can provide insight into the behavior of electrons in these systems. The CL of the Dirac equation is relevant for understanding the limits of measurement where we can gain insight into the fundamental limits of what we can measure in the quantum world. It also helps to connect with the macroscopic world, namely by understanding how quantum mechanics transitions to classical mechanics in the limit of large systems, we can better understand how quantum mechanics connects to the macroscopic world. We have what we call decoherence, which is where quantum systems tend to lose their coherence and become localized in a particular state. Understanding the mechanisms of decoherence can help us to design and control quantum systems, which is important for applications such as quantum computing. The CL is also important for the study of quantum chaos, which refers to the conduct of quantum systems that exhibit classical chaotic behavior in the classical regime. Understanding quantum chaos is important for a variety of applications, such as in the study of energy transport in complex systems. Overall, the study of the CL of the Dirac equation is important for both fundamental and practical reasons and has many connections to other areas of physics and mathematics.

We now summarize our results. We examined the effect of the phase-space noncommutativity on Ehrenfest's theorem through the form of velocity and Lorentz force, which have been found deformed and not converge and coincide with their classical versions. Knowing that the noncommutativity was implemented using both the ⋆product and three-dimensional Bopp-Shift. Note that in the limits of $\Theta \to 0$ and $\eta \to 0$, the extended deformed systems reduce to that of ordinary quantum mechanics, obtained and discussed in the literature. The results of this work can be used to extend the study with the other used techniques to implement the CL of other relativistic wave equations.

# Funding

This research received no external funding.

# Declarations conflict of interest

The author declares no conflict of interest.

# Reference

[1]  R. Alicki. Search for a border between classical and quantum worlds. Phys. Rev. A, 65(3) 2002, 034104(1-4).
[2]  N. Bohr. Über die Serienspektra der Elemente. Z. Physik, 2 1920, 423-469.
[3]  A.O. Bolivar. Classical limit of fermions in phase space. J. Math. Phys, 42(9) 2001, 4020-4030.
[4]  M. Chaichian, M. M. Sheikh-Jabbari, A. Tureanu. Hydrogen Atom Spectrum and the Lamb Shift in Noncommutative QED. Phys. Rev. Lett, 86(13) 2001, 2716-2719.
[5]  P. Ehrenfest, Bemerkung über die angenäherte Gültigkeit der klassischen Mechanik innerhalb der Quantenmechanik. Z. Physik, 45 1927 455-457.
[6]  G. Friesecke, M. Koppen. On the Ehrenfest theorem of quantum mechanics. J. Math. Phys, 50 2009, 082102(1-6).
[7]  D.M. Gingrich. Noncommutative geometry inspired black holes in higher dimensions at the LHC. J. High. Energy. Phys, 2010(05) 2010, 0-22.
[8]  L. Gouba. A comparative review of four formulations of noncommutative quantum mechanics. Int. J. Mod. Phys. A, 31(19) 2016, 1630025(1-15).
[9]  J.M. Gracia-Bondia. Notes on Quantum Gravity and Noncommutative Geometry: New Paths Towards Quantum Gravity. Springer, Berlin, Heidelberg, 2010.
[10] R.W. Greenberg, A. Klein, C. T. Li. Invariant tori and Heisenberg matrix mechanics: a new window on the quantum-classical correspondence. Phys. Rep, 264(1-5) 1996, 167-181.
[11] I. Haouam, L. Chetouani. The Foldy-Wouthuysen transformation of the Dirac equation in noncommutative phase-space. J. Mod. Phys, 9 2018, 2021-2034.
[12] I. Haouam. Foldy–Wouthuysen Transformation of Noncommutative Dirac Equation in the Presence of Minimal Uncertainty in Momentum. Few-Body. Syst, 64 2023, 9(2-14).
[13] I. Haouam. On the noncommutative geometry in quantum mechanics. J. Phys. Stud, 24(2) 2020, 2002(1-10).
[14] I. Haouam. Two-dimensional Pauli equation in noncommutative phase-space. Ukr. J. Phys, 66(9) 2021, 771-779.
[15] I. Haouam. On the Fisk-Tait equation for spin-3/2 fermions interacting with an external magnetic field in noncommutative space-time. J. Phys. Stud, 24(1) 2020, 1801(1-10).
[16] I. Haouam. Dirac oscillator in dynamical noncommutative space. Acta. Polytech, 61(6) 2021, 689-702.
[17] I. Haouam. Analytical solution of (2+1) dimensional Dirac equation in time-dependent noncommutative phase-space. Acta. Polytech, 60(2) 2020, 111-121.
[18] I. Haouam., H. Hassanabadi. Exact solution of (2+1)-dimensional noncommutative Pauli equation in a time-dependent background. Int. J. Theor. Phys, 61 2022, 215(2-13).
[19] I. Haouam, A. S. Alavi. Dynamical noncommutative graphene. Int. J. Mod. Phys, A 37(10) 2022, 2250054(1-18).
[20] I. Haouam. On the three-dimensional Pauli equation in noncommutative phase-space. Acta Polytech, 61(1) 2021, 230-241.
[21] I. Haouam. Continuity equation in presence of a non-local potential in non-commutative phase-space. Open. J. Microphys, 9(3) 2019, 15-28.
[22] I. Haouam. Solutions of Noncommutative Two-Dimensional Position–Dependent Mass Dirac Equation in the Presence of Rashba Spin-Orbit Interaction by Using the Nikiforov–Uvarov Method. Int. J. Theor. Phys, 62 2023, 111(1-25).
[23] I. Haouam. The non-relativistic limit of the DKP equation in non-commutative phase-space. Symmetry, 11 2019, 223, 1-15.
[24] A.A. Hnilo. Simple Explanation of the Classical Limit. Found. Phys, 49 2019, 1365-1371.
[25] G.K. Kay. Exact wave functions from classical orbits. II. The Coulomb, Morse, Rosen-Morse, and Eckart systems. Phys. Rev. A, 65(3) 2002, 032101(1-23).
[26] H. Krüger. Classical limit of real Dirac theory: Quantization of relativistic central field orbits. Found. Phys, 23 1993, 1265-1288.

[27] L.M. Liang, B.H. Wu. Quantum and classical exact solutions of the time-dependent driven generalized harmonic oscillator. Phys. Scr, 68(1) 2003, 41-44.
[28] L.M. Liang, J.Y. Sun. Quantum-classical correspondence of the relativistic equations. Ann. Phys, 314(1) 2004, 1-9.
[29] L.M. Liang, L.S. Shu, B. Yan. Quantum-classical correspondence of the Dirac equation with a scalar-like potential. Pramana - J Phys, 72 2009, 777-785.
[30] J. Madore. An introduction to noncommutative geometry. In: H. Gausterer, L. Pittner, H. Grosse, (eds) Geometry and Quantum Physics. Lecture Notes in Physics. Springer, Berlin, Heidelberg, vol 543 2000.
[31] A.J. Makowski. Exact classical limit of quantum mechanics: Central potentials and specific states. Phys. Rev. A, 65(3) 2002, 032103, 1-5.
[32] P. Martinetti. Beyond the standard model with noncommutative geometry, strolling towards quantum gravity. IOP Publishing, 634 2015, 012001.
[33] N. Seiberg, E. Witten. String theory and noncommutative geometry. J. High. Energy. Phys, 1999(9) 1999, 0-32.
[34] H. Spohn. Semiclassical limit of the Dirac equation and spin precession. Ann. Phys, 282(2) 2000, 420-431.
[35] R.J. Szabo. Quantum field theory on noncommutative spaces. Phys. Rep, 378(4) 2003, 207-299.
[36] P.A. Tipler, L. Ralph. A. Modern Physics (5ed.). W. H. Freeman and Company. pp. 160–161. ISBN 978-0-7167-7550-8, 2008.